# TUNEABLE MAGNETIC PROPERTIES OF CARBON-SHIELDED NIPT-NANOALLOYS

*Ahmed A. El-Gendy*[*,1,2,3,4], Silke Hampel[3], Bernd Büchner[3], Rüdiger Klingeler[#,1,5]*

[1]*Kirchhoff Institute of Physics, Heidelberg University, INF 227, 69120 Heidelberg, Germany*
[2]*Nanotechnology and Nanometrology Laboratory, National Institute for Standards, Giza 12211, Egypt*
[3]*Leibniz Institute for Solid State and Materials Research (IFW), D-01171 Dresden, Germany*
[4]*Department of Chemistry, Virginia Commonwealth University, Virginia 23284, USA*
[5]*Centre for Advanced Materials (CAM), Heidelberg University, INF 225, 69120 Heidelberg, Germany*

*Author Information*

*Corresponding Authors*

*Email: aelgendy@vcu.edu
[#]Email: klingeler@kip.uni-hd.de

*Author Contributions*

The manuscript was written through contributions of all authors. All authors have given approval to the final version of the manuscript.

**Abstract**

Spherical NiPt@C nanoalloys encapsulated in carbon shells are synthesized by means of high-pressure chemical vapour deposition. Upon variation of the synthesis parameters, both the alloy core composition and the particle size of the resulting spherical $Ni_xPt_{1-x}$@C nanocapsules can be controlled. The sublimation temperatures of the Ni- and Pt-precursors are found key to control the alloy composition and diameter. Depending on the synthesis parameters, the diameters of the cores are tuneable in the range of 3 - 15 nm while the carbon coatings for all conditions are 1 - 2 nm. The core particle size decreases linearly upon increasing either of the sublimation temperatures while the Ni and Pt content, respectively, are linearly increasing with the related precursor temperature. Accordingly, the magnetic properties of the nanoalloys, i.e. magnetization, remanent magnetization and critical field, are well controlled by the two sublimation temperatures. As compared to bulk NiPt, our data show an increase of Stoner enhancement by nanoscaling as ferromagnetism appears in the Pt rich NiPt nanoalloy which is not observed for the related bulk alloys. The magneto-crystalline anisotropy constants K are in the range of $0.3 - 4 \times 10^5$ J/m$^3$ underlining that NiPt@C is a stable, highly magnetic functional nanomaterial.

The promises of nanotechnology are driven by novel physical properties appearing upon reducing the dimensions of bulk materials. However, controlling and tailoring of nanomaterials with respect of size, composition, shape etc. becomes increasingly challenging on the nanoscale. This is in particular relevant for magnetic nanomaterials in which, at the border of evolving superparamagnetism, the magnetic properties and hence the feasibility for applications crucially depends on tiny changes of the shape, size, and composition of the materials. Such applications of magnetic nanomaterials concern, e.g., magnetic refrigeration [1], data storage [2], magnetic imaging [3], labelling [4], sensing [5], and cell separation [6,7], targeted drug delivery [8], and hyperthermia-based anti-cancer treatment [9,10]. Exploiting the promises of applied nanoscience hence demands production of uniform nanoparticles with controlled properties which is still one of the main challenges in this field. In addition, the use of nanoscaled functional materials, e.g., for electronic and biomedical applications, may require shielding of the materials by a protective coating. A promising way to provide a biocompatible and chemically stable shield appears to be the coating with carbon, thereby preventing the oxidation of the core material, promoting dispersity, and protecting the biological environment and the filling material from each other. [11,12] Regarding magnetic functionality, transition metal nanoparticles are ideally suited. Their synthesis routes have been considerably improved and have been extended to nanoalloys in the last few years by applying new organometallic precursors. [13]

Alloy nanoparticles provide additional ways to optimize magnetic properties of nanomaterials for particular applications. [14] In the case of Pt-alloys, the alloys are cost-efficient and exhibit favourable catalytic behaviour as compared to pure Pt nanoparticles. [15] In addition, mixing of Pt with Ni, Co, or Fe as alloys yields higher magnetocrystalline anisotropy [16,17] while the average ordered magnetic moment decreases with higher Pt content [18]. Here, we demonstrate how carbon-shielded NiPt nanoparticles (NiPt@C) can be progressively synthesized under composition- and size-control. Our results show that the sublimation temperatures of the Ni- and Pt-precursors are key to tune which is well understood in terms of homogeneous nucleation. The magnetic properties of the nanoalloys can be tuned accordingly so that, e.g., the superparamagnetic blocking temperature and the saturation magnetization can be tailored. Our straightforward approach of controlled synthesis hence opens a novel route to complex nanosized functional materials.

**Result and Discussion**

*Sample synthesis*

In previous studies, we have introduced the synthesis and the characterization of various magnetic nanoparticles such as Fe, Co, Ni, and NiRu using the high-pressure chemical vapour deposition technique (HPCVD) [19,20]. In the work at hand, HPCVD has been applied for the synthesis of carbon coated NiPt nanoalloys. Here, nickelocene and (trimethyl) methylcyclopentadienyl platinum (IV) precursors is used as starting materials for the nanoalloying constituents Ni and Pt, respectively. The advantage of using these types of precursors is that they sublimate at rather low temperatures thereby allowing large variations of the sublimation temperature in the experiments. The metal-organic precursors (weight ratio Ni:Pt = 2:1) are in two crucibles each placed in a separately thermostated chamber where they are sublimated to the gas phase. Then, the vapour is transferred to the CVD reactor by means of flowing argon gas (1400 sccm). The temperature and the pressure inside the CVD reactor can be varied in the range of 200 - 1000 °C and 5 - 40 bar, respectively. In the vapour-phase synthesis of nanoparticles, a supersaturated vapour is produced where the particles are formed while their re-evaporation is prevented. If the degree of supersaturation is sufficient, and the reaction / condensation kinetics permits, particles will nucleate homogeneously. In order to control the synthesis process, three main parameters have to be controlled. The first one is the temperature inside the CVD reactor which is adjusted and fixed at 900 °C, the optimum temperature for production of the spherical nanoparticles [19]. Secondly, the pressure inside the CVD reactor might play a crucial role. However, there was no effect on the size of NiRu@C nanoparticles [20], therefore here the pressure is fixed at 13 bar, the optimum value of for producing spherical nanoparticles [19]. Such high argon gas pressure yields frequent collisions with the precursor gas thereby both cooling the atoms and decreasing their diffusion rate out of source region. If the diffusion rate is not reduced sufficiently, supersaturation is not achieved and only individual atoms or very small clusters of atoms are deposited on the collecting surface. The third main parameter is the sublimation temperature $T_S$ of the precursors. It mainly affects the particle size by controlling the amount of sublimated starting materials entering the CVD reactor but also to some extend influences the alloy composition and the amount of carbon in the material. We hence performed a series of experiments where the temperatures of the two sublimation chambers were fixed at different temperature in the regime $T_S^{Pt}$ = 50 - 95 °C and $T_S^{Ni}$ = 95 - 170 °C for the Pt and Ni chamber, respectively, while all other reaction conditions remained constant. To summarize, the HPCVD technique allows to achieve and modulate a high degree of supersaturation in the gas phase which we used to grow nanosized NiPt alloys shielded by amorphous carbon shells

*Morphology and structure*

HRTEM images of HPCVD-grown NiPt@C nanoalloys at ($T_S^{Ni}$, $T_S^{Pt}$) = (95 °C, 95 °C), (110 °C, 95 °C), (130 °C, 60 °C), and (130 °C, 95 °C) are shown in Fig. 1. For all chosen combinations of sublimations temperatures in the two chambers, the HRTEM images proof the formation of core-shell nanoparticles. The thickness of the carbon shells amounts to 1-2 nm (cf. right column of Fig. 1). In addition, there is a clear variation of particle size upon changing the sublimation temperatures. From the TEM studies we extract average sizes of the NiPt cores of $13 \pm 5$, $9 \pm 3$, $6 \pm 2$, and $2 \pm 1$ nm for the different synthesis conditions as shown in Fig. 2 which values agree well to the analysis of the XRD data shown below. We emphasize a small size distribution for the samples prepared at high sublimation temperatures as shown in Fig. 2 which exhibit average diameters of $6 \pm 2$, and $2 \pm 1$ nm).

Our X-ray diffraction data confirm the formation of NiPt alloys in the core of the nanoparticles. The respective XRD patterns for materials synthesized under variation of the sublimation temperatures are displayed in Fig. 3. The diffraction peaks at $2\theta$ = 40°, 46°, 68° and 82° correspond to the main reflections of the NiPt fcc structure, i.e. (111), (200), (220) and (311) [21]. Fig. 3 indicates a small effect of the variation of synthesis conditions on the peak positions and a pronounced one on the peak width. The latter enables estimating the mean particle core crystalline sizes according to the Scherrer equation [22]. The analysis of the peak widths yields mean particles diameters of $15 \pm 7$, $10 \pm 5$, $7 \pm 4$, and $3 \pm 1$ nm for the samples (Fig. 3) which corroborates well with the results from the size determination by TEM (Fig. 2). The resulting core sizes of the materials obtained under different synthesis conditions are summarized in Fig. 4 (a). There is a clear correspondence of the mean particle size on the sublimation temperature. We find that the particle size decreases linearly upon increasing either of the sublimation temperatures $T_S^{Ni}$ or $T_S^{Pt}$ of the two precursors serving as Ni and Pt sources. This effect can be qualitatively explained by the fact that the temperatures of the sublimation chambers control the evaporated amount of the precursors entering the CVD reactor. Increasing the sublimation chamber temperatures hence increases the nucleation density and yields a higher degree of supersaturation in the reaction chamber and therefore homogeneous nucleation of smaller nanoparticles.

In addition to the size dependence, there are shifts of the peak positions in Fig. 3. Since the lattice parameters in NiPt nanoparticles depend on the actual composition, these changes indicate variations of the alloy composition. Extracting the lattice constants for each sample from the XRD data and comparing them to literature data on $Ni_xPt_{1-x}$ [23] enables determining the alloy composition (see Table 1). In addition, we have performed EDX studies, too, which confirm the compositions extracted from the lattice parameters. The results are displayed in Fig. 4b where a clear correspondence of the respective metal content in the material on the temperature in the Ni or Pt sublimation chambers is demonstrated. Quantitatively, we find again a linear dependence of the Ni and Pt content on the sublimation temperature of the Ni- or Pt-precursors, respectively, which again is associated with the fact that the temperature of the sublimation chambers directly controls the evaporated amount of the respective precursors entering the reactor zone.

*Magnetic Properties*

The magnetic properties of NiPt@C have been studied by means of static magnetization measurements. Fig. 5 presents the field dependencies of the magnetization M(H), at room temperature, i.e. the full hysteresis loops, in external magnetic fields up to 1 T. The main features are as following: For all materials, the data imply finite remanent magnetization $M_r$ and coercivity $H_C$. In addition, there is a tendency for saturation visible at higher fields. The quantitative analysis of the data reveals a strong dependence of the magnetization at B = 1 T on the sublimation temperatures: the higher the ratio $T_S^{Ni}$ : $T_S^{Pt}$ (i.e., the larger the difference between the temperatures) the higher values of M at B = 1 T are observed (Fig. 6). The data show that M(T = 300K, B = 1T) is affected by two key parameters: the magnetic properties of the alloy core material in general (which depends on the alloy composition) and the amount of magnetic material with respect to diamagnetic carbon [24]. The comparison with the corresponding data of pure $Ni_xPt_{1-x}$ bulk material with the same compositions *x* [23] hence enables to infer the actual carbon content in the materials as shown in Fig. 6. The relative amount of carbon monotonously increases upon increasing either sublimation temperature $T_S^{Ni}$ or $T_S^{Pt}$. Note, that the actual amount of carbon in the samples implies average interparticle distances of 8 - 20 nm between the particle centres. At these distances, only small dipolar interactions in the range of a few K are expected which can be neglected in our analysis of the room temperature magnetization.

Our finding of finite $M_r$ and $H_C$ at room temperature allows conclusions on the magnetic cores. (1) None of the samples under study is fully superparamagnetic at 300 K which is in stark contrast to the finding of a critical Ni-concentration of about 40 % in bulk samples below which no ferromagnetism occurs. (2) In all samples at least some of the particles have blocking temperatures above 300 K. The appearance of ferromagnetism for the nano-sized materials with high Pt content >60 % at hand may be associated with size effects on the electronic band structure. Indeed, ferromagnetism has been recently observed experimentally in carbon-coated platinum nanoparticles [25] and in alkanethiol-coated Pt nanoparticles [26]. Such behaviour may be explained by a further increase of the Stoner factor which is already significantly enhanced and close to ferromagnetism in bulk Pt. In Ref. [26], ferromagnetism in coated Pt nanoparticles is discussed in terms of both Stoner-like electronic band-magnetism and orbital ferromagnetism due to charge transfer the coating and Pt cores. This observation indicates a very large magnetic anisotropy which is expected if the large spin-orbit coupling in Pt is considered. The relevance of Pt for the observed large magnetic anisotropy is underlined by the fact that the critical field $H_C$ at room temperature monotonously decreases upon reducing the Pt content in the material down to 45 % (cf. Fig. 6a) while the particle diameters seem to play a minor role.

Further insight into the magnetic properties can be deduced from the temperature dependence of the magnetization displayed in Fig. 7. At 300 K, in all samples there is a difference between the zero-field cooled (ZFC) and the field cooled (FC) magnetization which excludes bare para- or superparamagnetism at room temperature. However, there are more features visible: (1) all curves show a kink in the ZFC magnetization curves at low temperature between 7 and 45 K (labelled by △ in Fig. 7). (2) The irreversibility temperature $T_{irr}$ (marked by ▲) below which the ZFC and FC curves depart from each other is larger than 400 K except for the lowest Pt content, i.e. $Ni_{0.6}Pt_{0.4}$. For this sample, there is a smeared blocking temperature-like feature at about 225 K (marked by ◇). This blocking temperature may be associated with the mean particle size in this material ($d \sim 10$ nm) while the experimentally observed particle size distribution (± 5 nm from the TEM analysis) suggests that the superparamagnetic blocking temperatures cover a large temperature range, i.e. up to $T_{irr}$.

The maxima of the ZFC curves in Fig. 7 indicate the blocking temperatures $T_B$ of 55, 8, 28, 224 K for samples prepared at $T_S^{Ni}$, $T_S^{Pt}$ = (110, 95), (130, 95), (130, 60), and (130, 50) °C, respectively. In Fig. 7d, there is another ZFC maximum at 18 K which confirms the existence of smaller particles as well as the relatively large size distribution in this sample. The blocking temperature dependence on the sublimation temperature shown in Fig. 7 implies that $T_B$ decreases linearly upon increasing of the sublimation temperature. The obtained particle sizes are linked to $T_B$ via the Neel-Brown equation $K_{eff}V = 25 \times$ Boltzmann constant $k_B\ T_B$ [27,28,29]. By assuming that the mean blocking temperature at the ZFC maxima is associated with the mean diameter as given above, the data hence yield magneto-crystalline anisotropy constants $K_{\text{eff}}$ of 0.3 - 4 x $10^5$ J/m$^3$ for the materials under study. These values are high compared to bulk Ni where $K_{\text{eff}}$ equals $4.5 \times 10^3$ J/m$^3$ [30] and to Ni nanoparticles with $K_{\text{eff}} \approx 1.3 \times 10^4$ J/m$^3$ [31] but are similar to $K_{\text{eff}} \approx 2.6 \times 10^5$ J/m$^3$ in nanoscaled nano-scaled NiRu@C [20]. The associated thermal energy barrier $K_{\text{eff}}V$ is on the order of 10 times the thermal energy $k_B T$ at room temperature which implies a good thermal stability which renders the nanoalloy promising for applications demanding high but not outstanding anisotropy such as cell labelling and separation or magnetic hyperthermia.

*Conclusions*

NiPt@C nanoalloys have been synthesized using the high pressure chemical vapour deposition technique and have been investigated with respect to their morphology and magnetic properties. Core/shell nanostructures of NiPt@C are formed with tuneable mean core size range from 3 to 15 nm as derived from HRTEM and XRD analysis. X-ray diffraction investigation reveals NiPt binary alloys only. Remarkably, the content of carbon in the deposited nanoalloy material as well as the actual Ni:Pt ratio linearly depend on the sublimation temperatures of the precursors. Also the produced particle size can be tuned by monitoring the sublimation temperature of each precursor. Both the composition and core size directly affect the magnetic properties. The magnetization curves show ferromagnetic-like behaviour at room temperature for all samples. The enhancement of the ferromagnetism of Pt by nanoscaling is evident as ferromagnetism appears in the Pt rich NiPt nanoalloy which is not observed for the related bulk alloys. This observation implies a further increase of Stoner enhancement which is large in bulk Pt by downscaling. Different values for the magnetization at B = 1 T are obtained by variation of chemical compositions of the NiPt cores but are strongly affected by different carbon contents in the deposited material, too. The thickness of the carbon shells is relatively constant. From the temperature dependent of the magnetization measurement, the ZFC curve suggests that the superparamagnetic blocking temperatures cover a large temperature range. The magneto-crystalline anisotropy constants K were calculated for all the samples to be in the range of 0.3 - 4 x $10^5$ J/m$^3$. The obtained data open root of optimising NiPt@C as a stable, highly magnetic functional nanomaterial.

*Experimental Methods*

HPCVD has been applied for the synthesis of carbon coated NiPt nanoalloys. The setup has been explained in detail before [19,20]. High-resolution transmission electron microscopy (HRTEM) was realised by means of a FEI Tecnai F 30 TEM with field emission gun at 300 kV. A Miniflex X-ray diffractometer (XRD) with Cu Kα radiation was used to identify the crystal structure. The magnetic field dependence of the magnetization at room temperature was measured by means of a MicroMag Model 2900 (Princeton Measurement Corp.) Alternating Gradient Magnetometer (AGM). The temperature dependence of the magnetization at constant magnetic field in the temperature range 5 to 400 K was studied in a Superconducting Quantum Interference Device (VSM-SQUID) Magnetometer from Quantum Design. For the zero-field cooled (field cooled) protocol, the samples were firstly cooled down to 5 K in zero magnetic field (in H = 100 Oe) and then the measurements were performed in a magnetic field of 100 Oe upon heating up to 400 K.

*Acknowledgements*

We thank G. Kreutzer and S. Pichl for technical support.

**Figures**

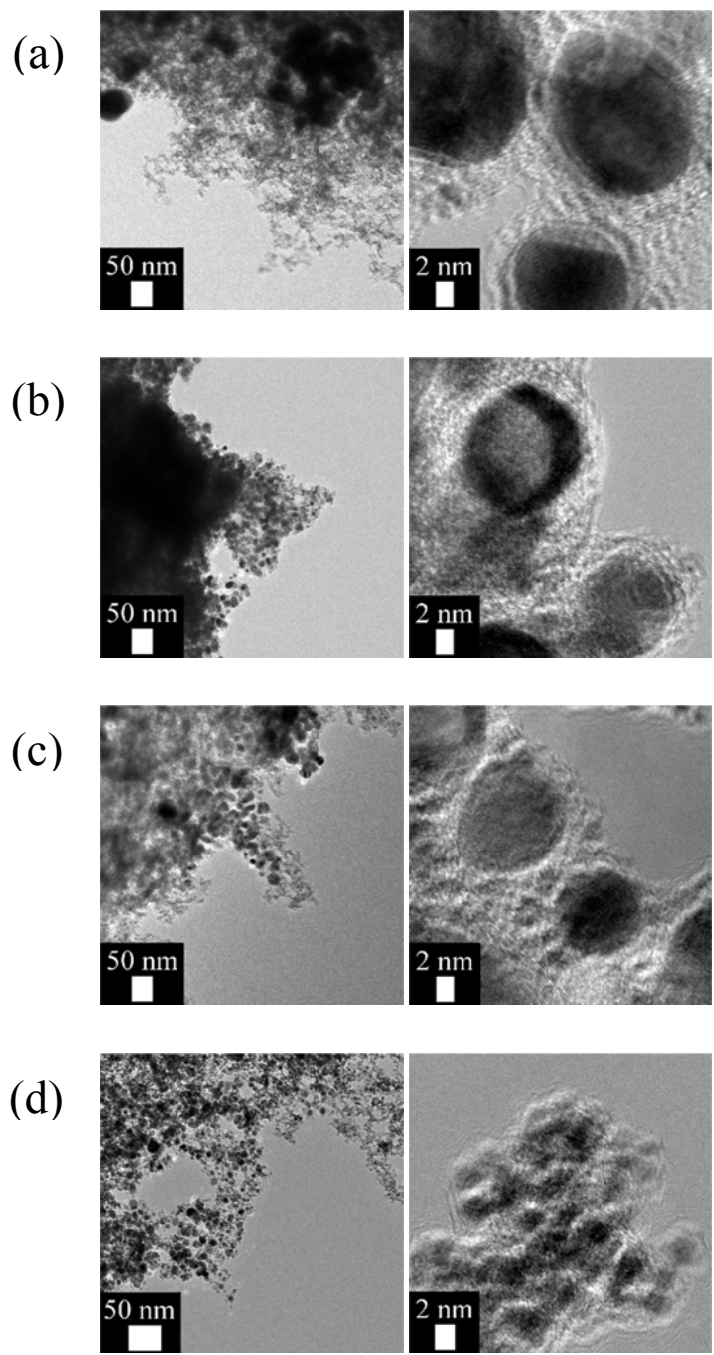

Figure 1: HR-TEM images of NiPt@C prepared at different temperatures in the sublimation chambers. Materials have been prepared at $(T_S^{Ni}, T_S^{Pt})$ = (95 °C, 95 °C) (a), (110 °C, 95 °C) (b), (130 °C, 60 °C) (c), and (130 °C, 95 °C) (d).

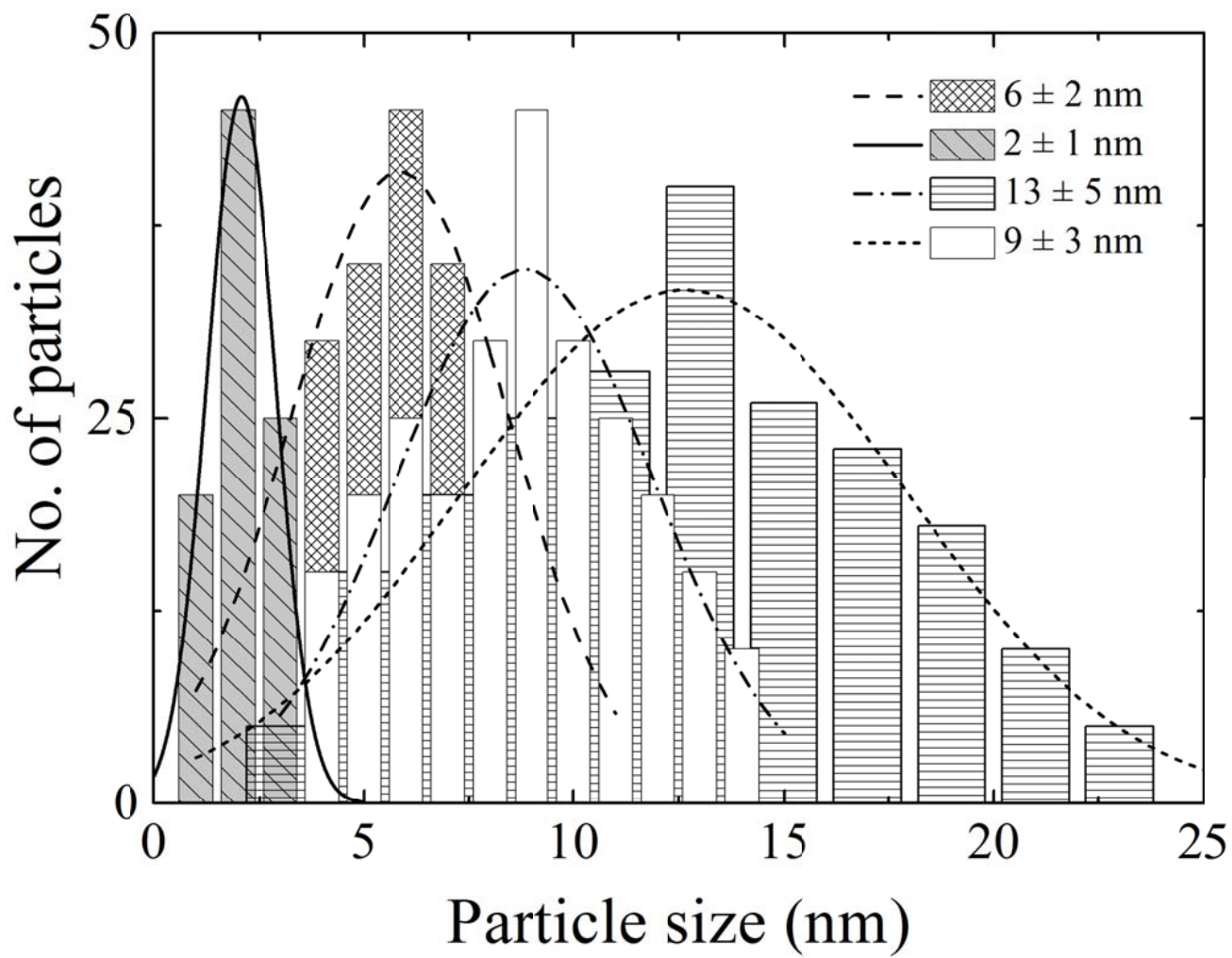

Figure 2: Size distribution obtained from TEM images for the samples shown in Fig.1 [row (c)], [row (d)], [row (a)] and [row (b)] respectively. Lines are Gaussian fits to the data.

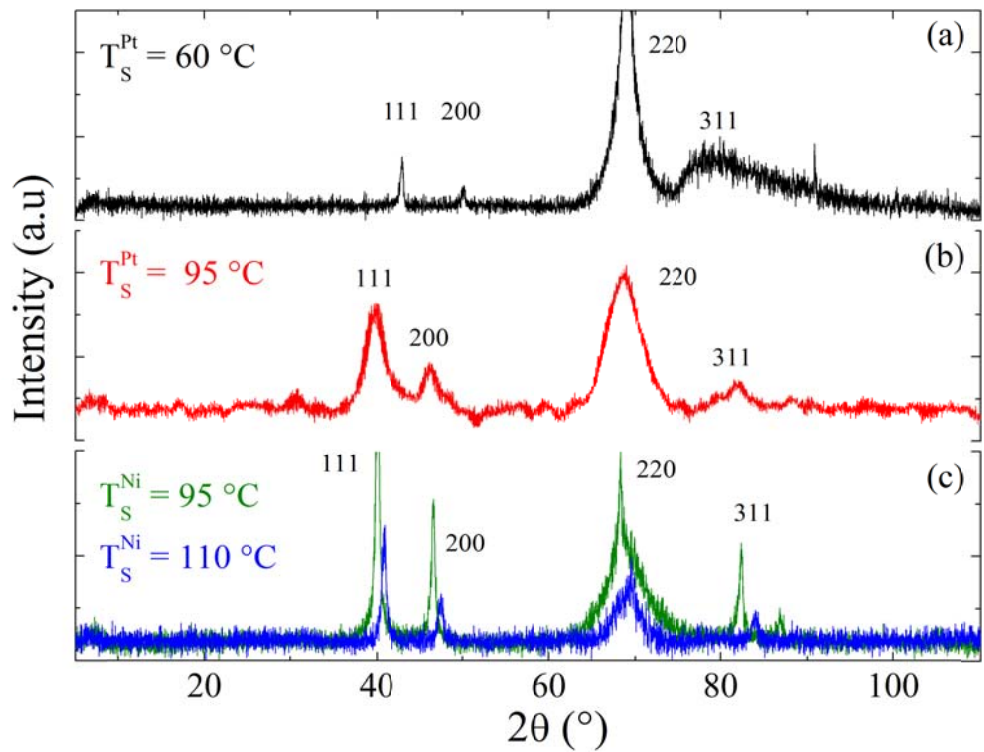

Figure 3: X-ray diffraction spectra of NiPt@C nanostructures for samples prepared at different sublimation temperatures of (a, b) the Pt-precursor at fixed $T_S^{Ni}$ = 130 °C and (c) Ni-precursor at fixed $T_S^{Pt}$ = 95 °C.

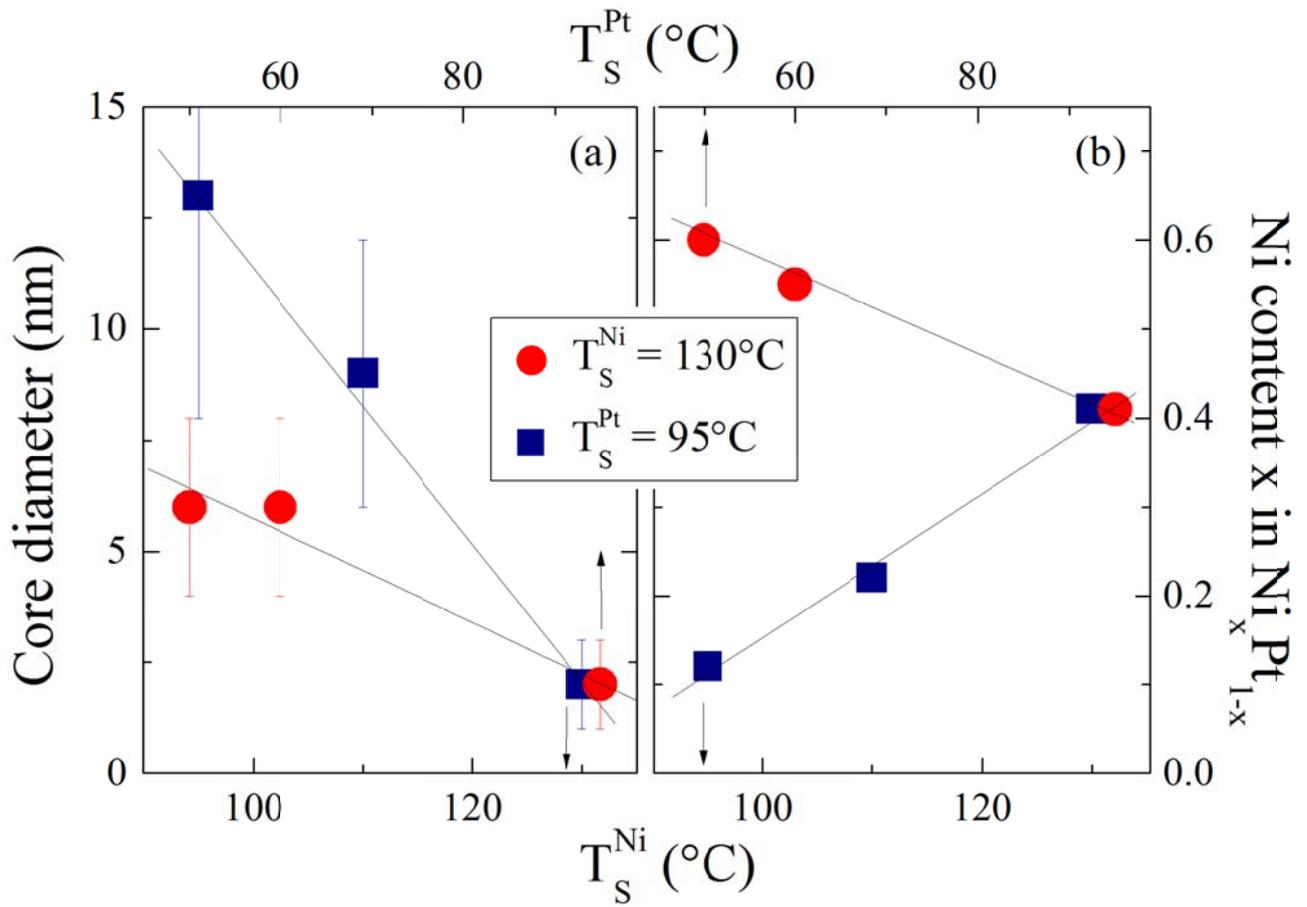

Figure 4: Particle core size (a) and alloy composition (b) in dependence on the sublimation temperature of the Pt-precursor $T_S^{Pt}$ at fixed $T_S^{Ni}$ = 130 °C (red circles, upper scale) and the Ni-precursor $T_S^{Ni}$ at fixed $T_S^{Pt}$ = 95 °C (blue squares, lower scale).

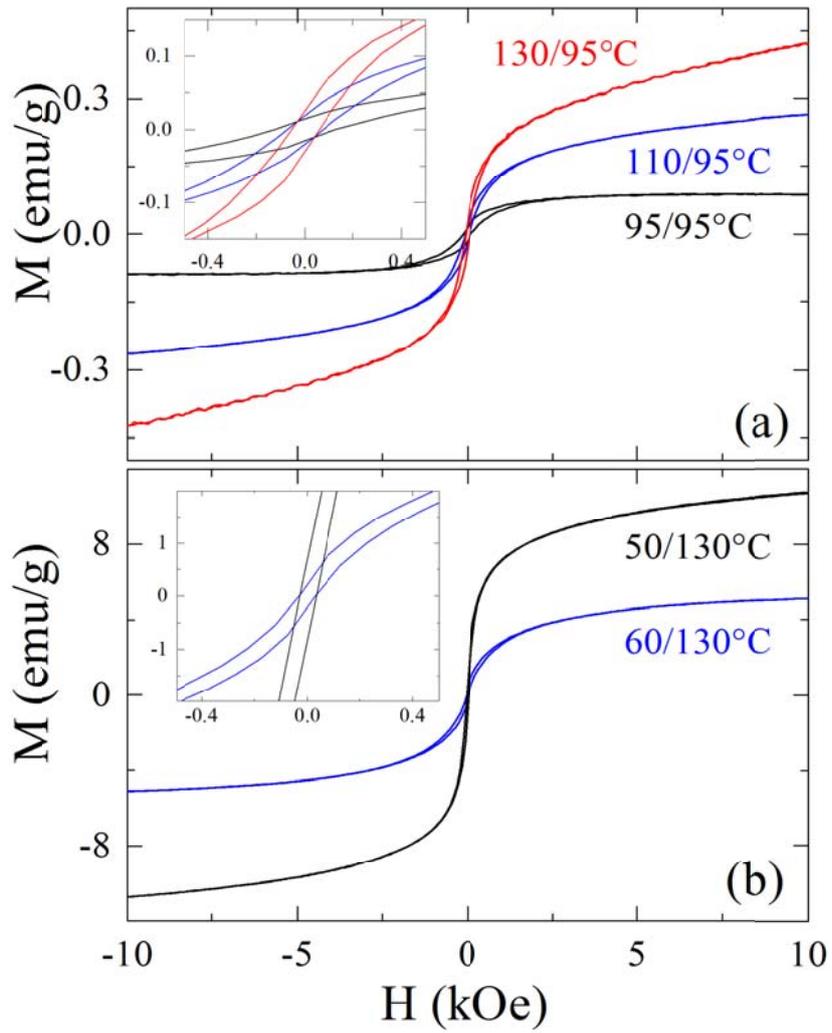

Figure 5: Magnetization loops at room temperature of NiPt@C nanoalloys synthesized at different sublimation temperatures (a) various $T_S^{Ni}$ with fixed $T_S^{Pt}$ = 95 °C, and (b) various $T_S^{Pt}$ with fixed $T_S^{Ni}$ = 130 °C. Insets show enlargements of main plots.

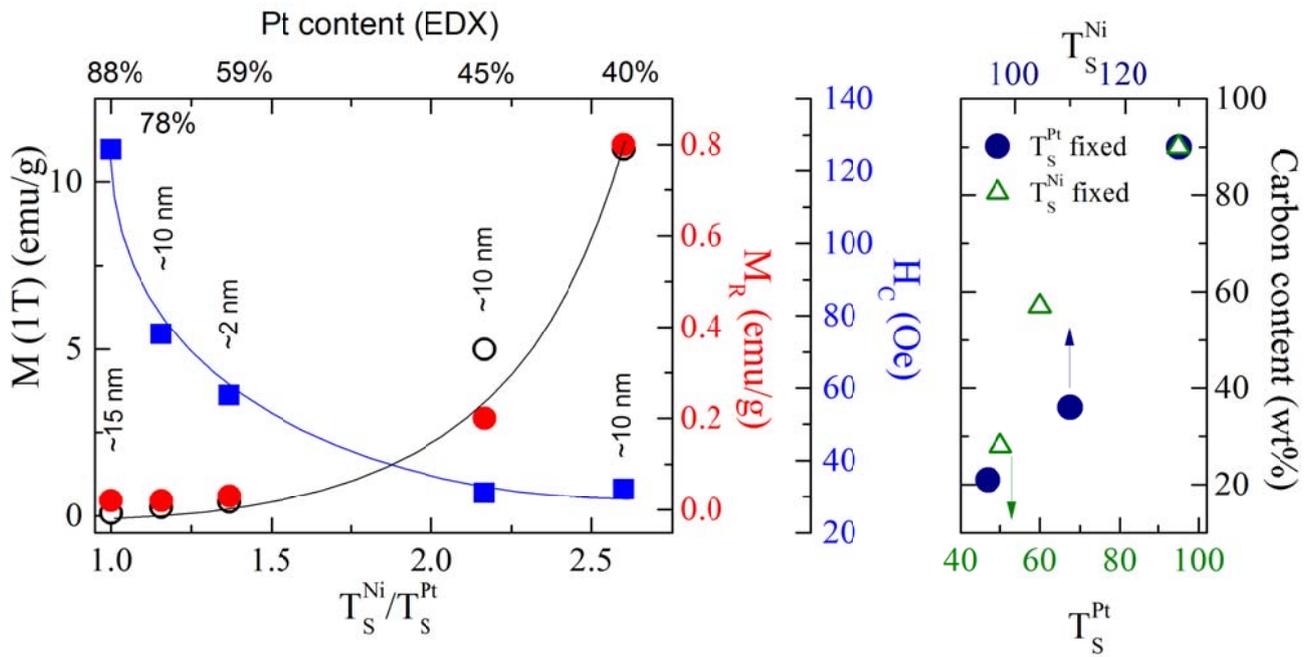

Figure 6: (a) Magnetization at B = 1 T (open black circles), remanent magnetization $M_R$ (filled red circles) and critical field $H_C$ (blue squares) at T = 300 K vs. ratio $T_S^{Ni}/T_S^{Pt}$ of sublimation temperatures of the two precursors. (b) Carbon content as extracted from the analysis of the magnetization data upon variation of each sublimation temperature separately.

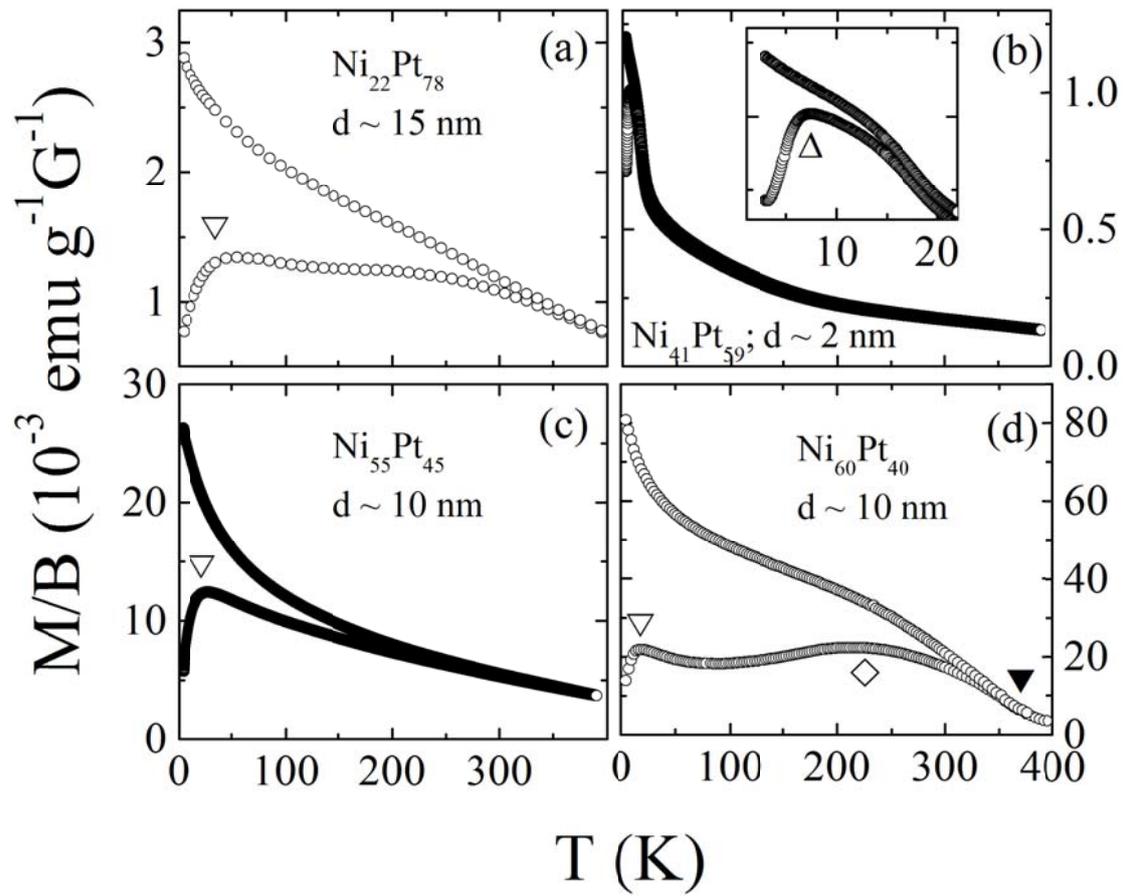

Figure 7: Temperature dependence of the magnetization of NiPt@C prepared at $T_S^{Ni} / T_S^{Pt}$ = (a) 110 °C/ 95°C, (b) 130 °C/95 °C, (c) 130 °C/60 °C, and (d) 130 °C/50 °C. Symbols are discussed in the text.

Table 1: Materials parameters of differently synthesized samples magnetization, EDX, XRD, and TEM studies in addition to bulk data which are extrapolated from Refs. [23] and [18].

| $T_s^{Ni}$ (°C) | $T_s^{Pt}$ (°C) | $x_{Ni}:x_{Pt}$ (EDX) | bulk $T_C$ (K) Ref. [23] | Average moment ($\mu_B$/atom) Ref. [18] | $H_C$ (Oe) | $M_s$ (emu/g) | $M_r$ (emu/g) | $d_{TEM}$ (nm) | $d_{XRD}$ (nm) | $T_B$ (K) | $a_{XRD}$ (nm) | $x_{Ni}:x_{Pt}$ (XRD) |
|---|---|---|---|---|---|---|---|---|---|---|---|---|
| 95 | 95 | 12:88 | non-magn. | - | 126 ± 2 | 0.09 | 0.02 | 13 ± 6 | 15 ± 7 | ≥400 | 0.3825 ± 0.001 | 28:72 |
| 110 | 95 | 22:78 | non-magn. | - | 75 ± 2 | 0.26 | 0.02 | 9 ± 4 | 10 ± 5 | 55, (214) | 0.380 ± 0.01 | 34:66 |
| 130 | 95 | 41:59 | ~10 | ~0.05 | 58 ± 2 | 0.42 | 0.03 | 2 ± 1 | 3 ± 1 | 7.5 | 0.376 ± 0.002 | 50:50 |
| 130 | 60 | 55:45 | ~100 | ~0.25 | 31 ± 2 | 5 | 0.2 | 9 ± 4 | 10 ± 5 | 28 | 0.374 ± 0.01 | 55:45 |
| 130 | 50 | 60:40 | ~161 | 0.3 | 32 ± 2 | 11 | 0.8 | 9 ± 4 | 10 ± 5 | 18, 224 | 0.374 ± 0.01 | 55:45 |